\documentclass{optica-article}
\usepackage{listings}
\journal{opticajournal} 
\usepackage{float}
\articletype{Research Article}
\usepackage{ulem}
\usepackage{xcolor}

\usepackage{color}

\begin{document}

\year=2025

\begin{center}
\fbox{
\begin{minipage}{0.95\linewidth}
\small

\textbf{Attribution Statement}

© 2025 Optica Publishing Group. One print or electronic copy may be
made for personal use only. Systematic reproduction and distribution,
duplication of any material in this paper for a fee or for commercial
purposes, or modifications of the content of this paper are prohibited.

This is the accepted manuscript of the article:

G. Elmas, I. A. Litvin, P. Kohl, and J. Nötzel,
``Modeling and Analysis of Phase Instability in Photonic Processor,''
\emph{Applied Optics}, 2025.

DOI: \url{https://doi.org/10.1364/AO.560370}

\end{minipage}
}
\end{center}

\title{Modeling and Analysis of Phase Instability in Photonic Processor}

\author{Gökhan Elmas\authormark{1,*}, Igor A Litvin \authormark{1}, Paul Kohl \authormark{1} and Janis Nötzel\authormark{1,**}}

\address{\authormark{1}Emmy-Noether Group Theoretical Quantum Systems Design, Technical University of Munich, Munich, Germany}
\email{\authormark{*}gokhan.elmas@tum.de} 
\email{\authormark{**}janis.noetzel@tum.de}

\begin{abstract}
Achieving both reconfigurability and stable output signals is a critical challenge in the development of integrated photonic circuits for large-scale optical quantum information processing. This has led to the creation of multimode photonic processors, also known as reconfigurable multimode interferometers, which have wide-ranging applications in quantum and classical information processing. However, maintaining phase stability in multi-port input signals remains a significant hurdle, particularly due to the phase instabilities introduced by active cooling systems and temperature drifts in the photonic processor. In this study, we propose theoretical models to simulate phase instability in photonic processors and validate them against experimental results. Two distinct modeling approaches were employed: a Brownian random walk and phase reconstruction based on experimentally observed oscillating harmonics. Additionally, we verified and applied our model to a
specific application for input phase correction using self-feedback control within the photonic
processor.
\end{abstract}


\section{Introduction}
As conventional electronic computing approaches its limits in terms of speed, energy efficiency, and parallelism, there is growing interest in transitioning to optical, or photonic, computing. Photonic processors, which utilize light rather than electrons for computations, promise significant advantages in areas such as quantum computing\cite{Harris2016,Qiang2018,Silverstone2016}, secure communication\cite{Paraiso2021}, signal processing\cite{Wang2018}, and machine learning\cite{Cheng2020}. However, early implementations of photonic computing relied on bulk optical components, which, while demonstrating the potential of optics in computation, faced serious challenges in terms of scalability, stability, and integration\cite{OBrien2003}. These bulky systems were difficult to miniaturize and lacked the practicality required for widespread use. To address these issues, silicon photonics has emerged as a transformative advancement, allowing the integration of optical components onto compact, chip-scale platforms. This approach not only retains the benefits of photonic computing but also provides the scalability, manufacturability, and cost-effectiveness needed for large-scale applications in quantum computing—for tasks such as secure quantum information processing—and in neural network implementations, where parallel optical signal processing can enhance computation speed and efficiency. Silicon photonic processors, particularly those employing rectangular meshes of reconfigurable Mach-Zehnder Interferometers (MZIs), have proven to be highly versatile for implementing complex linear operations and unitary transformations\cite{Clements2016}. Despite these advancements, challenges such as environmental noise and mechanical instabilities still need to be addressed for full-scale adoption.

The use of light for computational tasks leverages the wave phenomenon of interference between input signals. This allows tunable optical components, such as beam splitters and phase shifters, to be configured for algorithmic purposes \cite{Arrazola2021,Zhong2020}. However, stable and reliable use of interference as a computational model depends on the stability of relative phases between input modes, as dephasing can disrupt the output intensity distributions. One of most important source of dephasing is mechanical vibrations in fiber-optic cables connecting the laser source to the silicon photonic device. The geometric alterations in fiber-optic cables caused by mechanical vibrations also lead to changes in the refractive index during transmission, directly affecting phase stability  in optical systems \cite{Kiiveri2022,Svarc2023}. In practical settings, vibrations induce microbends along the fiber, which, although small, lead to localized distortions in the fiber geometry and result in fluctuations in the effective refractive index. These index variations can disrupt the conditions for total internal reflection, altering the propagation path of light and introducing phase noise into the transmitted signal. Such phase instabilities significantly degrade computational accuracy, underscoring the need for effective noise mitigation strategies to preserve the integrity of photonic computing.

Numerous attempts have been made to achieve robust phase and intensity correction in photonic systems. For example, \cite{Svarc2023} demonstrated sub-0.1$^\circ$ phase locking of a single-photon interferometer by employing a classical reference signal at a different wavelength combined with adaptive detection thresholds to compensate for intensity fluctuations. In contrast, \cite{Zhong2020} achieved significant stability in a large-mode interferometer by integrating active laser phase locking with passive techniques to minimize long-term drift. Similarly, \cite{Smith2009} demonstrated phase-controlled quantum interference in UV-written integrated photonic circuits using a thermo-optic phase shifter, where local heating via a NiCr electrode enabled precise and tunable phase stabilization within Mach-Zehnder interferometers. These studies underscore the critical importance of developing accurate phase noise models to simulate such processes, thereby enhancing the overall performance and reliability of photonic setups.

In this work, we present an experimental investigation of phase instability in a photonic processor, comparing it to simulation results that utilize two distinct approaches: a Brownian random walk and phase reconstruction derived from experimentally obtained oscillating harmonics, ensuring comprehensive accounting of all spectral features of the input signals.  We examine the effects of this noise on the stability and performance of the processor, particularly in tasks requiring precise phase control. By quantifying the impact of phase noise on computational accuracy and assessing its implications for system stability, we aim to contribute to the understanding and mitigation of environmental disturbances in photonic processing. This work highlights the importance of addressing phase noise to ensure the reliability and scalability of silicon photonic platforms in computational applications. Additionally, we verified and applied our model to a specific application for input phase correction using self-feedback control within the photonic processor.

\section{Brownian motion model for phase instability}

The stability of phase in optical systems is crucial for maintaining computational accuracy in photonic processors. However, mechanical vibrations in fiber-optic cables introduce random fluctuations in key parameters that affect phase stability. These fluctuations primarily manifest as variations in the refractive index \( n \) and in the propagation path length \( L \), both of which contribute to cumulative phase noise in the transmitted signal. This random noise, resulting from changes in \( n \) and perturbations in light propagation, can interfere with the precision of computations. Consequently, a comprehensive analysis of noise generated by fiber vibrations is essential, with a focus on its stochastic influence on refractive index variations and propagation path shifts. Such insights are fundamental to developing robust compensation strategies that can stabilize optical computations and mitigate phase-related noise disruptions.

The phase \( \phi \) of the monochromatic coherent light traveling through a medium with fluctuations in the refractive index or propagating distance is given by 
\( \phi(t) = 2\pi L_n(t)/{\lambda} + \omega t \), where \( L_n(t)\) is the effective propagating length, \( n \) is the refractive index of the medium, \( \lambda \) is the wavelength of the light in a vacuum, \( \omega = 2\pi f \) is the angular frequency of the light with \( f \) being its frequency, and \( t \) represents time, describing both the spatial and temporal evolution of the light's phase. To simplify the equations, we will assume single-wavelength monochromatic light and omit the \(\omega t\) term. We see that the minor variations in \( L \) and \( n \) will induce corresponding shifts in phase. Practical systems experience such variations due to mechanical vibrations affecting the optical fibers, leading to stochastic fluctuations in \( L_n(t) \). These phase fluctuations can be effectively modeled as a random-walk process where the changes in \( L_n(t) \) follow a Brownian motion model \cite{Kuschnerov2010, Seimetz2009}. Brownian motion aptly describes the cumulative behavior of these mechanical vibrations, as it encapsulates the random, independent, and incremental nature of the perturbations in the optical path. Over time, the cumulative effects of these random perturbations increase the uncertainty in phase, thereby resulting in growing phase noise that can interfere with computational processes in photonic systems.

Modeling the phase noise as a Brownian process allows us to quantify and simulate the stochastic nature of these fluctuations. By understanding the statistical properties of phase noise generated by mechanical disturbances, it becomes possible to design more effective strategies for compensating phase-related noise, thereby enhancing the stability and performance of photonic computing architectures. 

Let us consider the evolution of the phase fluctuation  due to optical fiber vibration (which is linearly proportional to the effective propagation length fluctuation \( \sim  L_n(t)\)) as a Brownian motion process, denoted by \(\phi(t)\), in which changes over non-overlapping time intervals are independent and normally distributed.
Specifically, we assume that

\begin{equation}
\phi(t + \Delta t) - \phi(t) \sim \mathcal{N}(0, \operatorname{var}({\phi}(\Delta t)), \label{eq:normal}
\end{equation}

\noindent where \( \mathcal{N} \) is the normal distribution with a mean of zero, and \(\operatorname{var}({\phi}(\Delta t))\) is the variance of the phase change over a specific time period \(\Delta t\).

Eq. \eqref{eq:normal} characterizes the statistical behavior of the stochastic process \(\phi(t)\) over an infinitesimal time increment \(\Delta t\). Specifically, it indicates that the change in \(\phi(t)\) follows a normal distribution with a mean of zero and a variance proportional to the length of the time increment.
In the case of Brownian motion the phase variance can be presented as:

\begin{equation}
\operatorname{var}({\phi}(\Delta t)) = {\sigma^2 \Delta t},
\end{equation}
where \(\sigma^2\) represents the phase variance value . Such a behavior is characteristic of Brownian motion, which is governed by independent increments that follow a normal distribution.

To derive the distribution of \(\phi(t)\) at a fixed time \(t\), one must consider the cumulative effect of all preceding increments up to that point. By summing these independent increments, the process \(\phi(t)\) can be expressed as the total of these stochastic variations. Since each increment contributes a variance of \(\sigma^2 \Delta t\), the overall variance over a time period \(t\) accumulates to \(\sigma^2 t\).

This relationship illustrates that the phase change variance increases linearly with time. Such a linear growth in phase noise reflects the cumulative effect of stochastic perturbations in refractive index and path length, emphasizing the need for effective noise compensation mechanisms in optical systems to maintain phase stability over time.

\section{Experimental setup and simulation methodology}

The photonic processor used in this study is an 8-mode configurable mesh of Mach-Zehnder Interferometers (MZIs) and tunable phase shifters (TPSs), integrated on-chip and arranged in a rectangular configuration. By adjusting the phase shifts within each configurable MZI, we implement complex, unitary transformations across the processor, facilitating programmable operations for diverse photonic computing tasks. The processor’s mesh architecture enables reconfigurability and scalability, essential for applications in optical signal processing, quantum information, and machine learning.

Each unit cell of the photonic processor comprising a tunable MZI and TPS is allowing precise control over phase and amplitude modulation and described by the following unitary matrix

\begin{equation}   
U{(\phi,\theta)} = \frac{1}{2}
\begin{pmatrix} 1 - e^{-i\theta} & -i (e^{-i\theta} + 1) \\ -i (e^{-i\theta} + 1) e^{-i\phi} & -(1 - e^{-i\theta}) e^{-i\phi} \end{pmatrix}{\qquad \theta\in[0,2\pi),\ \phi\in[0,2\pi)}
\end{equation}

\noindent where $\theta$ and $\phi$ correspond to implemented configurations of the MZI and TPS respectively.

\begin{figure}[h!]
    \centering
    \includegraphics[width=0.99\textwidth]{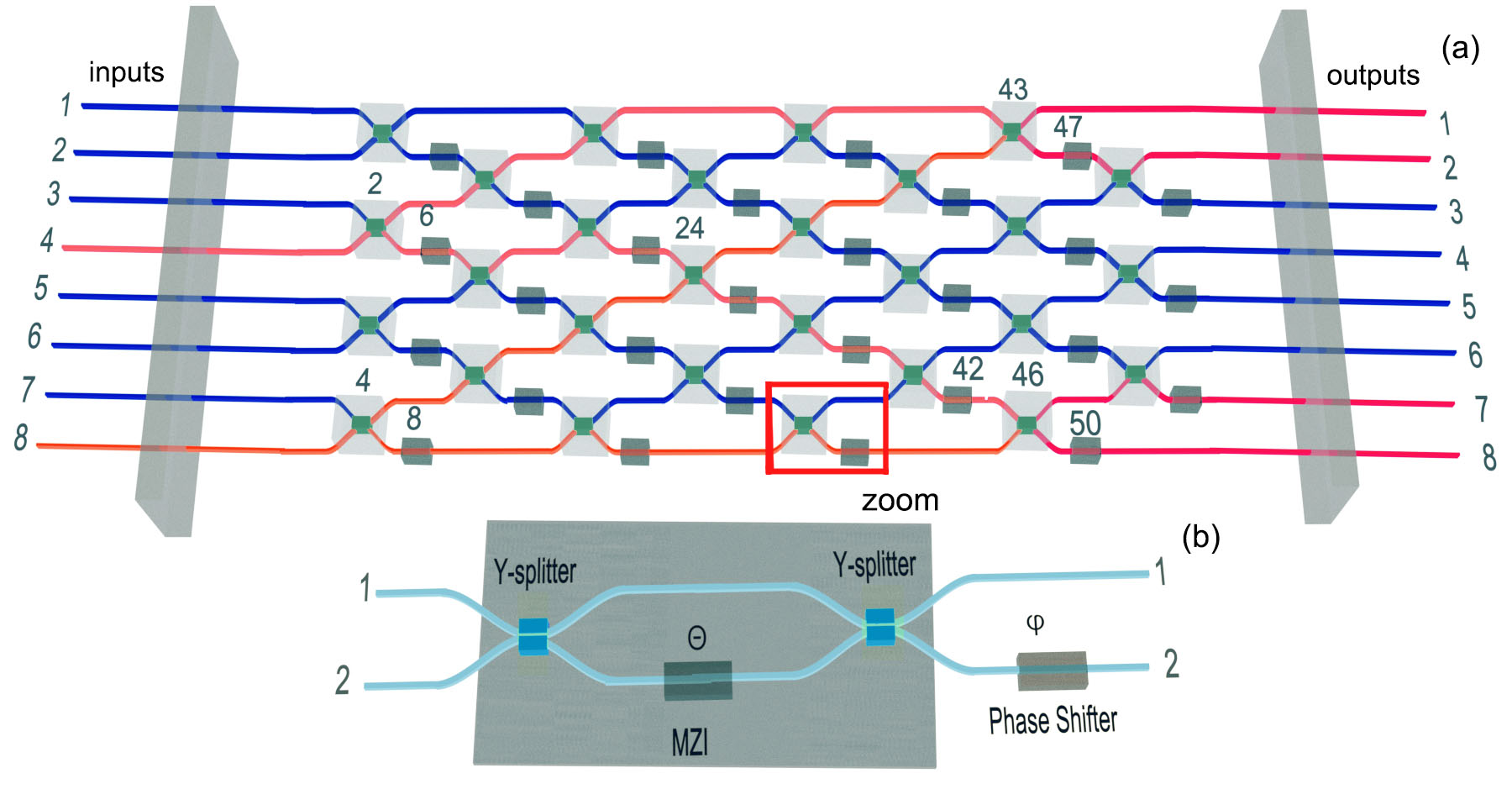}
    \caption{(a) Schematic of the photonic processor featuring an 8-mode mesh of MZIs and TPSs. The red and orange lines indicate the two input signal paths. (b) A zoomed-in view of the highlighted unit cell (red rectangle in (a)), which consists of an MZI and an external phase shifter, illustrating the basic building block of the processor.}
    \label{fig:chip}
\end{figure}

To determine the relative phase difference between two input signals produced by undesirable shaking of the connecting optical fibers, we implement a transformation from a two-mode input state to a four-mode output state. This is achieved by configuring the MZIs  and TPSs within the photonic processor, as depicted in Fig.~\ref{fig:chip}. In the unit cell comprising MZI[2] and TPS[6], the optical signal introduced through Input 4 is split into two components of equal amplitude. Similarly, the unit cell comprising MZI[4] and TPS[8] is utilized to equally split the light introduced through Input 8.

The equally split light components are then merged in two unit cells: one consisting of MZI[43] and TPS[47], and the other consisting of MZI[46] and TPS[50]. In both cases, the MZIs are configured to $\pi/2$ and works as the 50-50 beam splitter. By tuning TPS[42] to $\pi/2$, an additional phase is introduced to one of the inputs. This results in outputs at ports 7 and 8 depending on the cosine function, while outputs at ports 1 and 2 depend on the sine function in the absence of the additional phase. This configuration allows us to extract phase information across the entire complex plane.

The extension from a two-input state to a four-output state is crucial for extracting the relative phase between the input beams using the tangent function. By expressing the output powers in terms of sine and cosine, we can directly compute the tangent of the relative phase across the complex plane. This is achieved through the phase differences that arise from the sine and cosine terms, enabling phase extraction without the typical restrictions of the interval \( [-\pi, \pi] \) and truncation of the extracted phase. The reparameterization of the output powers allows us to map the full range of phase values, making it possible to accurately track the relative phases. By unwrapping the extracted phases, we are able to capture phase variations that extend beyond conventional periodicity, which is essential for handling experimental data where phase values can exceed \( [-2\pi, 2\pi] \) bounds. 

For our particular case of the experimental setup and 8 mode photonic processor (see Fig. 1), the input signals with  the same  amplitudes \( E_0 \) at inputs 4 and 8, denoted  as \( E_{\text{in4}} \) and \( E_{\text{in8}} \), respectively, can be described as follows:

\begin{subequations}
\begin{align}
    E_{\text{in4}}(t) &= E_0 e^{i\phi_4(t)}, \label{eq:in4} \\
    E_{\text{in8}}(t) &= E_0 e^{i\phi_8(t)}. \label{eq:in8}
\end{align}
\end{subequations}
where \( \phi_{4,8}(t) \) is the oscillating phases of the input signals.

By configuring the unit cells in a rectangular mesh according to the specific parametrization described previously and taking into account the transformation matrix of every unit cell (see Eq. 3), the relationship between the input and output amplitudes is given by:

\begin{subequations}
\begin{align}
    E_{\text{out1,out2}}(t) &= \pm \frac{i}{2} E_{\text{in4}}(t) + \frac{1}{2} E_{\text{in8}}(t), \label{eq:out1_out2} \\
    E_{\text{out7,out8}}(t) &= -\frac{i}{2} E_{\text{in4}}(t) \mp \frac{i}{2} E_{\text{in8}}(t). \label{eq:out7_out8}
\end{align}
\end{subequations}
where \( E_{\text{out1}} \),\( E_{\text{out2}} \),\( E_{\text{out7}} \) and \( E_{\text{out8}} \) represent the output amplitudes at ports 1,2,7 and 8, respectively. 

The respective output powers \( P_{\text{out1}} \), \( P_{\text{out2}} \), \( P_{\text{out7}} \), and \( P_{\text{out8}} \) are:

\begin{subequations}
\begin{align}
    P_{\text{out1,out2}}(t) &= \left(  \pm \frac{i}{2} E_{\text{in4}}(t) + \frac{1}{2} E_{\text{in8}}(t) \right) 
    \left(  \mp \frac{i}{2} E_{\text{in4}}(t)^* + \frac{1}{2} E_{\text{in8}}(t)^* \right) \notag \\
    &= \frac{1}{4} \left( |E_{\text{in4}}(t)|^2 + |E_{\text{in8}}(t)|^2 
    \pm \operatorname{Im}\big(E_{\text{in4}}(t) E_{\text{in8}}(t)^*\big) \right) \notag \\
    &= \frac{E_0}{2} \left( 1 \pm \sin\big(\Delta\phi_{\text{in}}(t)\big) \right), \label{eq:P_out1_out2} \\
    P_{\text{out7,out8}}(t) &= \left(   - \frac{i}{2} E_{\text{in4}}(t) \mp \frac{i}{2} E_{\text{in8}}(t) \right) 
    \left(   \frac{i}{2} E_{\text{in4}}(t)^* \pm \frac{i}{2} E_{\text{in8}}(t)^* \right) \notag \\
    &= \frac{1}{4} \left( |E_{\text{in4}}(t)|^2 + |E_{\text{in8}}(t)|^2 
    \pm \operatorname{Re}\big(E_{\text{in4}}(t) E_{\text{in8}}(t)^*\big) \right) \notag \\
    &= \frac{E_0}{2} \left( 1 \pm \cos\big(\Delta\phi_{\text{in}}(t)\big) \right), \label{eq:P_out7_out8}
\end{align}
\end{subequations}
where \( \Delta\phi_{\text{in}}(t) \) represents the relative and undesirable phase difference induced by fiber vibrating between the input signals \( \phi_4(t) \) and \( \phi_8(t) \).

 Here, \( P_{\text{out1}} \), \( P_{\text{out2}} \), \( P_{\text{out7}} \), and \( P_{\text{out8}} \) are the optical powers measured at the output ports 1, 2, 7, and 8, respectively.

After normalizing the output power, we can obtain required phase difference, \( \Delta\phi_{\text{in}} \), between input signals 4 and 8:

\begin{subequations}
\begin{align}
    \Delta\phi_{\text{in}}(t) &= \arcsin\big(2P_{\text{out1}}(t) - 1\big), \label{eq:arcsin1} \\
    \Delta\phi_{\text{in}}(t) &= \arccos\big(1 - 2P_{\text{out8}}(t)\big). \label{eq:arccos2}
\end{align}
\end{subequations}

Based on Eq.~\eqref{eq:arcsin1} and Eq.~\eqref{eq:arccos2}, we can obtain two phase values that are offset by a \(\pi/2\) phase difference. This phase difference allows us to reconstruct the continuous phase behavior without being constrained by the boundaries of the interval \( [-\pi, \pi] \), where phase wrapping occurs, leading to discontinuities in the reconstructed phase.   

For example, when one solution from Eq.~\eqref{eq:arcsin1} reaches the \( [-\pi, \pi] \) boundary, we can switch to the solution from Eq.~\eqref{eq:arccos2} to avoid phase truncation. By alternating between the two equations, it becomes possible to incrementally extract the true phase values, ensuring a seamless and continuous phase reconstruction. 

\begin{figure}[H]
   \centering
    \includegraphics[width=0.99\textwidth]{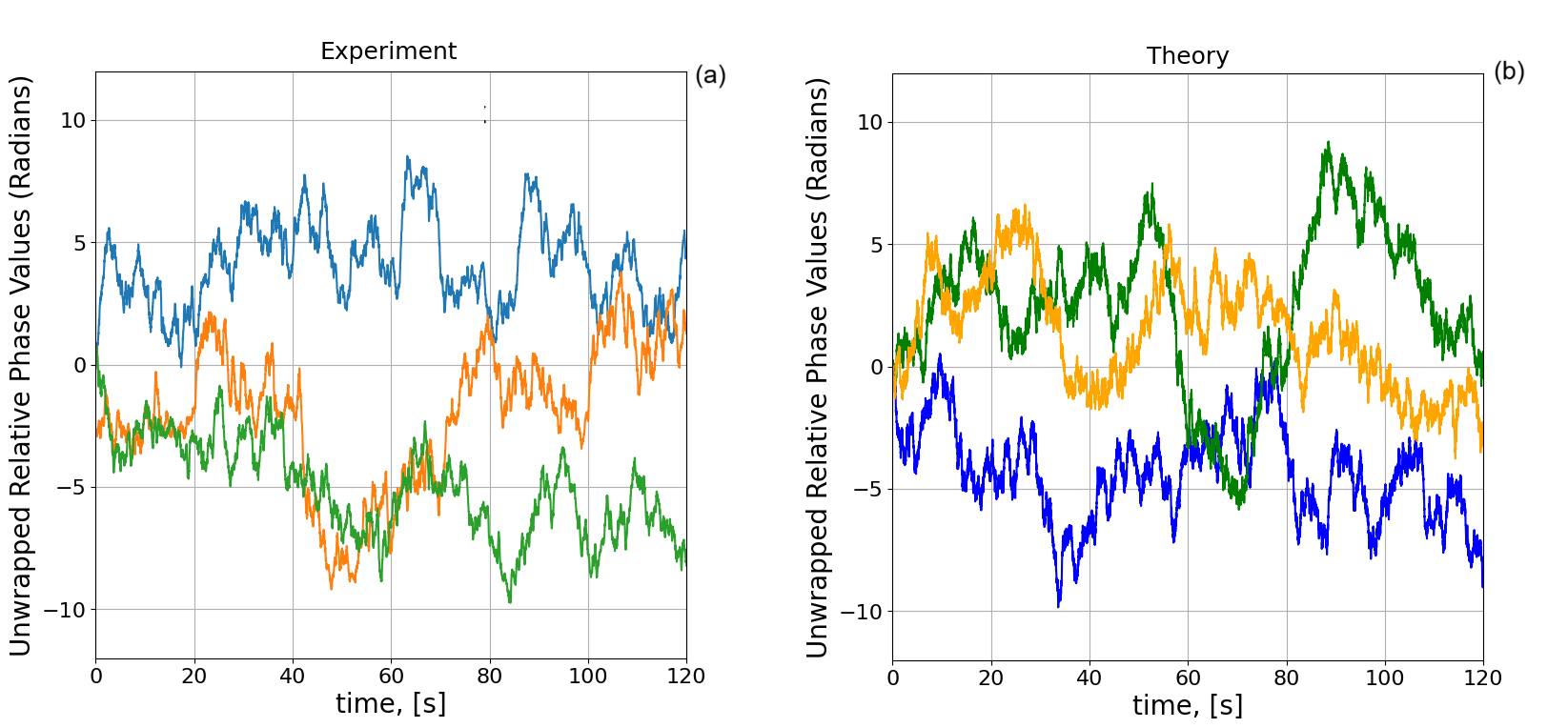}\caption{Comparison of the unwrapped extracted relative phase values: (a) extracted from three experiments, and (b) extracted from three simulated experiments. Representative phase traces obtained from simulation (based on a stochastic Brownian model) and from experiment. The simulation captures the overall statistical behavior of phase fluctuations, while the experimental trace reflects a typical measurement under our operating conditions.}
    \label{fig:unwrapped}
\end{figure}

The unwrapped extracted phase values obtained from three experiments and simulations are presented in Fig.~\ref{fig:unwrapped}, which illustrates example phase traces from both the simulation and experimental measurements.

Following this phase extraction, we applied  Fourier Transform (FT) to the phase values to analyze the frequency components of phase fluctuations. FT provides insight into the oscillatory behavior of $\Delta\phi(t)$, allowing us to identify the dominant frequencies and phase relationships in the signal.

In our practical applications, $\Delta\phi(t)$ is sampled at discrete time intervals, resulting in a finite set of data points. For such discrete data, the Discrete Fourier Transform (DFT) is used to approximate the FT. Given a discrete sequence $\Delta\phi(n)$, where $n = 0, 1, \ldots, N-1$ represents $N$ equally spaced time samples, the DFT is defined as:
\begin{equation}
    F[k] = \sum_{n=0}^{N-1} \Delta\phi(n) e^{-i \frac{2 \pi}{N} k n}
\end{equation}
where  $F[k]$ is the $k$-th frequency component in the discrete spectrum, $N$ is the number of samples,  $k = 0, 1, \ldots, N-1$ indexes the discrete frequency components.

The DFT provides a discrete approximation of the continuous FT and produces a frequency spectrum $F[k]$ suitable for analyzing the sampled data. The amplitude $|F[k]|$ indicates the strength of each frequency component in $\Delta\phi(n)$, while $\arg(F[k])$ provides phase information.

By systematically applying these data processing steps, we ensured that the noise characteristics and behavior of the photonic processor were effectively captured, enabling a robust assessment of the system’s stability and response to environmental disturbances. This processing framework, designed for both real and simulated data, thus provided a comprehensive basis for evaluating the impact of phase noise on photonic computation.

\section{Comparison of model simulation and the experimental results}

In our experimental setup, the output powers  were collected using a high-precision analog-to-digital converter (ADC) card integrated with a Raspberry Pi control unit for streamlined data acquisition.  The experiment yielded approximately 125 readings per second per channel and this dataset provides a basis for subsequent analysis.

In alignment with the data processing methodology discussed earlier, the normalized output powers were used to extract the relative phase information between the two input ports. According to the theoretical model (see Eq. 2), the presence of Brownian motion predicts that the extracted phase values should follow a normal distribution, with the variance exhibiting a linear relationship with elapsed time. 

In the simulation, each input mode to the processor was independently subjected to noise, reflecting real-world conditions where spatial modes experience varying phase disturbances over time. The simulation generated eight independent random walk sequences  for the eight spatial modes of the processor. Each sequence represented phase noise over time, with a predefined variance dictating the magnitude of fluctuations at each time step. Each random walk was constructed by calculating a cumulative sum of normally distributed increments scaled by the square root of the time step, ensuring adherence to Brownian motion properties (see Eq. 2). This approach captures the stochastic nature of phase disturbances observed in optical fibers connecting the laser to the photonic chip, caused by unpredictable mechanical and thermal instabilities.

The noisy output signals were processed using the same methodology as the experimental data, including normalization and Fourier analysis. This consistent approach facilitated a direct comparison between simulated and experimental results, offering insights into the photonic processor’s sensitivity to environmental noise and the potential efficacy of noise mitigation strategies.

We conducted thirty-five experiments, each following the scheme described in Eq. 7 (a,b). Each experiment lasted 300 seconds, during which we continuously monitored the statistical distribution of the relative phase values. The primary objective of these trials was to analyze the behavior of the extracted phase values across multiple experiments and compare these observations with theoretical predictions. 

\begin{figure}[H]
   \centering
    \includegraphics[width=0.99\textwidth]{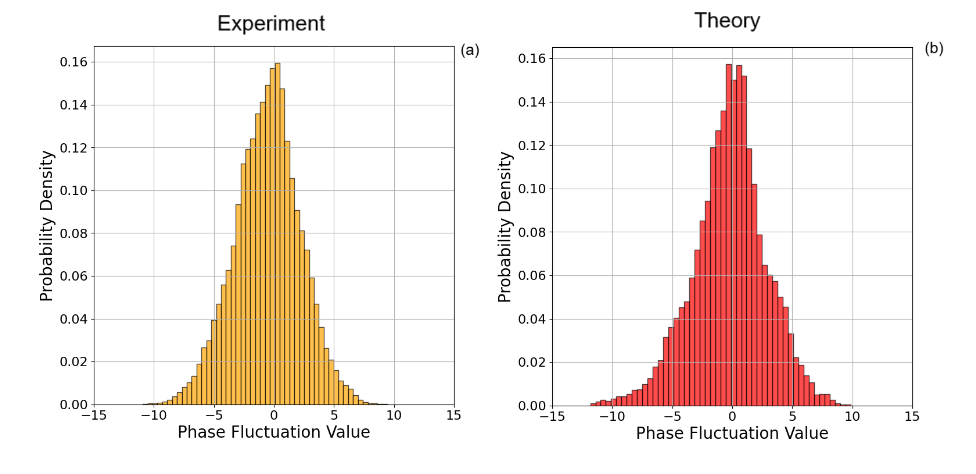}\caption{Comparison of the distribution of relative phase values: (a) extracted from thirty-five experiments, and (b) extracted from thirty-five simulated experiments with $\sigma=0.25 \ [sec]^{-1/2}$. The discrepancy between the simulated and experimental data is approximately $9\%$.}
    \label{fig:CombinedHistogram}
\end{figure}

The resulting distribution, shown in Fig.~\ref{fig:CombinedHistogram} (a), reveals key insights into the statistical characteristics of the phase values. It demonstrates consistency across trials and aligns closely with the theoretical framework outlined in Eq. 7 (a, b). This agreement underscores the robustness of the proposed approach in capturing the underlying statistical properties of phase dynamics.

To complement the experimental findings, simulations were conducted to analyze the distribution of relative phase values under controlled noise conditions. In these simulations, the standard deviation was set to $\sigma = 0.25 \ [sec]^{-1/2}$, with the duration and number of simulated experiments mirroring those of the actual experiments. The results, shown in Fig.~\ref{fig:CombinedHistogram} (b), reveal the statistical characteristics of phase fluctuations in the simulated scenarios, enabling direct comparison with the experiments. The close agreement between the experimental and simulated distributions demonstrates the validity of the theoretical model and the reliability of the experimental setup.

\begin{figure}[h!]
    \centering
    \includegraphics[width=0.99\textwidth]{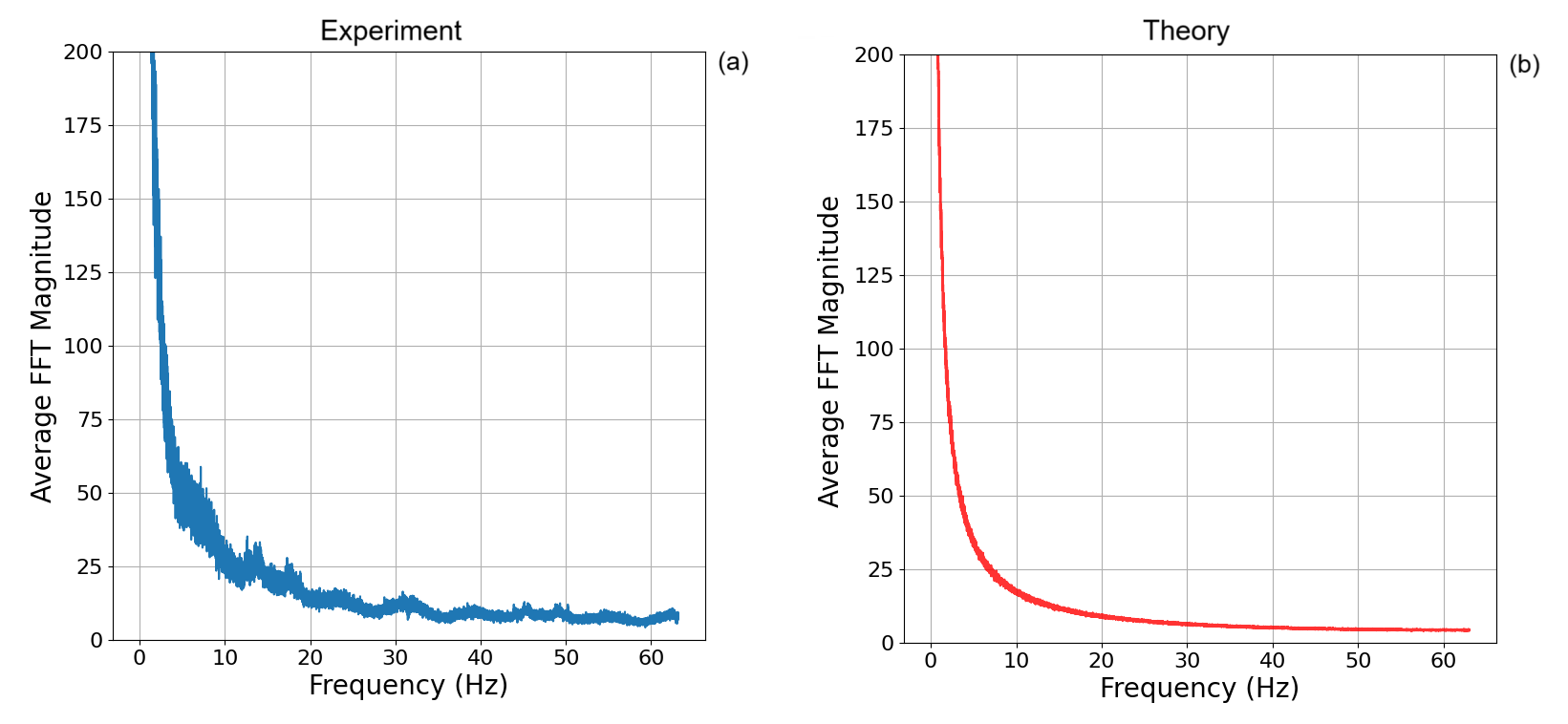}
    \caption{Comparison of Fourier transforms of the extracted phase values: (a) averaged across 35 experimental trials, and (b) derived from simulations under software controlled noise conditions. The discrepancy between the simulated and experimental data is approximately $12\%$.}
    \label{fig:CombinedFFT}
\end{figure}

To analyze the frequency characteristics of the extracted phase values, we performed DFT on the data from both experimental and simulated trials. The resulting frequency spectra, shown in Fig.~\ref{fig:CombinedFFT}, provide insights into the noise contributions within the dataset. DFT of the experimental data, averaged across 35 trials, reveals a characteristic spectrum dominated by Brownian noise effects.

The results demonstrate a strong correspondence between the theoretical model, experimental observations, and simulated outcomes. The statistical distribution of relative phase values, shown in Fig.~\ref{fig:CombinedHistogram}, and the frequency spectra derived through Fourier analysis, presented in Fig.~\ref{fig:CombinedFFT}, exhibit good consistency across experimental and simulated datasets. The observed noise characteristics, identified as Brownian noise, further reinforce the robustness of the theoretical framework in capturing the stochastic dynamics of the phase values. This close alignment affirms the reliability of the experimental and simulation methodologies employed. Together, these findings provide a comprehensive validation of the proposed model and its applicability to understanding phase dynamics in complex systems.

\section{Phase reconstruction by the experimentally obtained oscillation spectrum}
In the case of precise reconstruction of the phase difference oscillation of the input signals in photonic processor we can apply the Inverse Fourier Transform (IFT) to the experimentally obtained spectrum. Such reconstructed phase difference will take into account all spectral features of the oscillating signal and can be used for more precise investigation.
After obtaining  $F[k]$ (see Eq. 8), IFT is used to reconstruct the original time-domain function. This step is essential to verify the completeness of the spectral representation and to prepare $\Delta\phi(n)$ for subsequent simulations.

In our case of discrete data, the Inverse Discrete Fourier Transform (IDFT) is used to reconstruct $\Delta\phi(n)$ from $F[k]$:
\begin{equation}
    \Delta\phi(n) = \frac{1}{N} \sum_{k=0}^{N-1} F[k] \, e^{i \frac{2 \pi}{N} k n}
\end{equation}
where $\Delta\phi(n)$ is the reconstructed time-domain signal, $F[k]$ is the discrete frequency spectrum obtained from the DFT.

\begin{figure}[h!]
    \centering
    \includegraphics[width=0.59\textwidth]{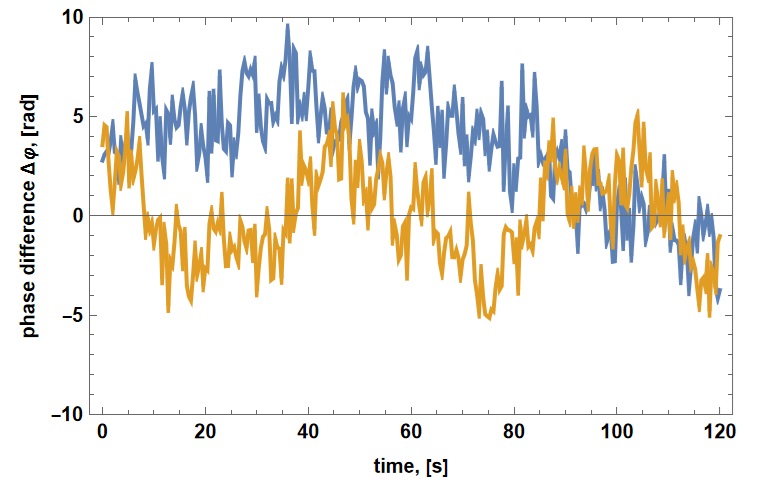}
    \caption{Examples of the reconstructed phase difference oscillations between two optical fibers. The reconstruction was performed using the IDFT on the experimentally obtained oscillation spectrum. The figure highlights how the full spectral features, including contributions from both Brownian motion and additional correlated noise (e.g., thermal drift, mechanical resonances), are captured to accurately restore the original phase behavior.}
    \label{fig:my_figure}
\end{figure}

The IDFT sums all frequency components $F[k]$, scaled by their respective amplitudes and phase shifts, to accurately reconstruct the original oscillatory behavior of $\Delta\phi(t)$. This reconstructed function can then be used in simulations and further analysis.

This theoretical framework enables the analysis and reconstruction of a time-dependent phase oscillation function. By computing the FT, we obtain a frequency-domain representation (power spectrum) that reveals the underlying frequency components of $\Delta\phi(t)$. Applying the IFT then allows us to accurately restore the original function, confirming the accuracy of the spectral representation and preparing $\Delta\phi(t)$ for subsequent applications in simulations.

In the Fig.~\ref{fig:my_figure} we have presented two example of the reconstructing phase difference oscillation which are based on the experimentally obtained  spectrum of $\Delta\phi(t)$.

This comprehensive phase reconstruction approach offers significant benefits in accurately capturing the dynamics of phase noise. By utilizing the full spectrum of observed oscillations, this method not only reproduces the stochastic fluctuations described by the Brownian model but also includes contributions from other correlated noise sources such as thermal drift and mechanical resonances. This detailed spectral analysis is critical for developing robust phase stabilization techniques, thereby improving the performance and reliability of photonic processors in complex computational applications.

\section{Implementation}

In this section, we implement the theoretical results obtained to simulate the processes and output signal behavior of the photonic processor, comparing the simulations with experimental data.

Fig. 2(a) illustrates the impact of phase instability in the input signals of the photonic processor. This instability arises primarily due to airflow generated by air-cooling fans, which causes vibrations in the connecting optical fiber. These vibrations result in unstable interference within the photonic processor (PP) chip. To address this issue, a Feedback Control (FBC) mechanism is implemented, designed to stabilize the relative phase difference between input signals and maintain consistent interference within the PP chip \cite{Litvin2024}.

The FBC scheme aims to mitigate phase instability between PP inputs using one of the PP phase shifters to correct the phase difference between the input signals and to improve the interference stability in the chip \cite{Litvin2024}.

\begin{figure}[H]
    \centering
    \includegraphics[width=0.9\textwidth]{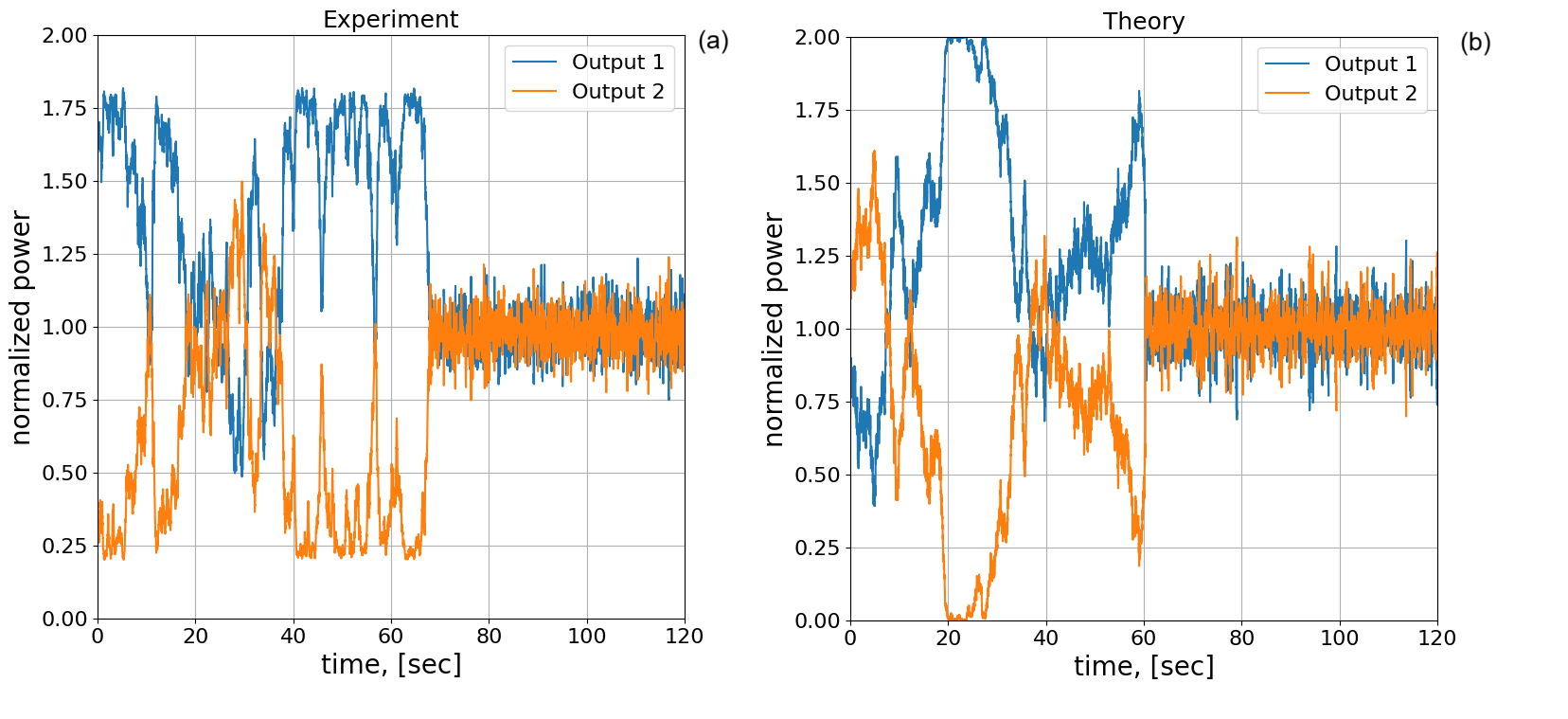}
    \caption{The comparison of phase stabilization for the two-input signal FBC scheme in the photonic processor is shown in the experimentally obtained data (a) and the simulated data (b). The FBC was activated after 60 seconds (off before 60 seconds and on thereafter).  In both cases we have used 38 Hz FBC correction rates and the simulation was generated with $\sigma=0.15 \ [sec]^{-1/2}$.}
    \label{fig:phasestabilization}
\end{figure}

Fig. 6 presents the experimental results of the FBC implementation alongside simulations of a similar FBC scheme, incorporating phase fluctuations modeled on Brownian motion. The experimental and simulated results demonstrate a high degree of consistency. In both cases, the FBC mechanism was activated after 60 seconds (remaining off prior to this point). The simulations used a standard deviation of $\sigma = 0.15 \ [sec]^{-1/2}$ for the phase fluctuations, and both cases we have used 38 Hz FBC correction rates which was limited by our current experimental setup \cite{Litvin2024}. The behavior observed in the experimental data closely aligns with the simulated outcomes, further validating the efficacy of the phase instability simulation approach described in this work.

Environmental vibrations can induce mechanical perturbations in both fiber-optic cables and the photonic chip, leading to fluctuations in optical path lengths and relative phase shifts. In fibers, vibrations cause microbending and localized distortions that alter the effective refractive index, while on-chip vibrations may induce strain and modify the waveguide geometry via the photoelastic effect. In our design, these deleterious effects are mitigated through a combination of passive and active strategies. First, the chip and fiber-optic interfaces are mechanically isolated using vibration-damping mounts and careful packaging techniques to reduce the transmission of environmental vibrations. Second, an active phase stabilization scheme based on self-feedback control (FBC) is implemented. This scheme continuously monitors the output interference pattern and dynamically adjusts one or more phase shifters to correct for any phase drift caused by vibrations.

\section{Conclusion}

In this work, we present an experimental investigation of phase instability in a photonic processor, comparing it with simulation results that model the noise using two approaches: a Brownian random walk and phase reconstruction based on experimentally obtained oscillating harmonics, ensuring that all spectral features of the input signals are accurately accounted for. We investigate the effects of phase noise on the stability and performance of the photonic processor, with a particular focus on tasks requiring precise phase control.

The results reveal a strong correspondence between the theoretical model, experimental observations, and simulated outcomes. The statistical distribution of relative phase values and the frequency spectra derived through Fourier analysis exhibit high consistency across experimental and simulated datasets. The noise characteristics, identified as Brownian noise, further validate the robustness of the theoretical framework in capturing the stochastic dynamics of phase values. This close alignment affirms the reliability of the experimental and simulation methodologies employed in this study.

To achieve precise reconstruction of the phase difference oscillations of the input signals in the photonic processor, we applied Fourier transform analysis to the experimentally obtained spectrum. This approach enabled us to reconstruct the oscillatory behavior of phase instability, incorporating all spectral features of the oscillating signal. The reconstructed phase difference serves as a valuable tool for more detailed investigations and simulations of phase instability in the photonic processor.

Furthermore, we validated and applied our model to a practical application for input phase correction using FBC within the photonic processor. The experimental data demonstrated FBC behavior that closely aligned with simulated outcomes, further confirming the effectiveness of the phase instability simulation approach presented in this work. The strong agreement between experimental and simulated distributions validates the theoretical model and underscores the reliability of the experimental setup.

By quantifying the impact of phase noise on computational accuracy and evaluating its implications for system stability, this work contributes to a deeper understanding of environmental disturbances in photonic processing and offers strategies for their mitigation. These findings underscore the critical importance of addressing phase instability to ensure the reliability and scalability of silicon photonic platforms for computational applications and demonstrate its applicability to understanding and mitigating phase dynamics in complex photonic systems.

The proposed model and its corresponding experimental validation offer valuable insights for a range of applications requiring high-precision phase control. In particular, these results can be used to optimize photonic processors for quantum computing, optical signal processing, and neuromorphic computing, where maintaining stable interference is crucial. Moreover, the simulation framework can serve as a design tool to inform future device fabrication and calibration strategies in advanced silicon photonic platforms.

\section*{Acknowledgement}

This work was financed by the DFG via grant NO 1129/2-1 and by the Federal Ministry of Education and Research of Germany in the programme of 'Souverän. Digital. Vernetzt.' Joint project 6G-life, project identification number: 16KISK002 and via grants 16KISQ077 and 16KISR026.

\section*{Disclosures}
The author declares no conflict of interest.

\section*{Data availability}
Data underlying the results presented in this paper are not publicly available at this time but may be obtained from the authors upon reasonable request.

\end{document}